\documentclass[12pt,preprint]{aastex}

\usepackage{color}

\slugcomment{Submitted to the \apjl}

\shorttitle{Asymmetric Deflagration}
\shortauthors{Calder et~al.}

\begin{document}

\title{Type Ia Supernovae: An Asymmetric Deflagration Model}

\author{
A. C. Calder\altaffilmark{1,2},
T. Plewa\altaffilmark{1,2,3},
N. Vladimirova\altaffilmark{1,2},
D. Q. Lamb\altaffilmark{1,2,4},	and
J. W. Truran\altaffilmark{1,2,4}
}

\altaffiltext{1}{Center	for Astrophysical Thermonuclear	Flashes,
   The University of Chicago,
   Chicago, IL 60637}
\altaffiltext{2}{Department of Astronomy \& Astrophysics,
   The University of Chicago,
   Chicago, IL 60637}
\altaffiltext{3}{Nicolaus Copernicus Astronomical Center,
   Bartycka 18,
   00716 Warsaw, Poland}
\altaffiltext{4}{Enrico	Fermi Institute,
   The University of Chicago,
   Chicago, IL 60637}

\begin{abstract}

We present the first high-resolution three-dimensional simulations of
the deflagration phase of Type Ia supernovae that treat	the entire
massive	white dwarf. We	report the results of simulations in which
ignition of the	nuclear	burning	occurs slightly	off-center.  The
subsequent evolution of	the nuclear burning is surprisingly asymmetric
with a growing bubble of hot ash rapidly rising	to the stellar
surface.  Upon reaching	the surface, the mass of burned	material is
$\approx 0.075 M_\sun$ and the kinetic energy is $4.3 \times
10^{49}$~ergs.	The velocity of	the top	of the rising bubble
approaches 8000~km\,s$^{-1}$.  The amount of the asymmetry found in the
model offers a natural explanation for the observed diversity in Type
Ia supernovae.	Our study strongly disfavors the classic 
central-ignition pure deflagration scenario by showing that the result
is highly sensitive to details of the initial conditions.

\end{abstract}


\section{Introduction}

Type Ia	supernovae are one class of stellar explosions noted for their
extreme	brightness and distinguished by	the lack of hydrogen in	their
spectra. These events are thought to originate in close	binary systems
in which the primary component is a degenerate C-O rich	white
dwarf~\citep{hoyle+60} gaining mass accreted from a low-mass main
sequence companion.  A thermonuclear explosion is believed to be
initiated in the core of the white dwarf when its mass approaches the
Chandrasekhar limit~\citep{whelan+73,livio00}. The exact nature	of
the ignition and subsequent evolution, however,	is the subject of
ongoing	research.

Regardless of the exact	nature of the explosion, the nuclear energy
released by the	decay of $^{56}$Ni and $^{56}$Co makes Type Ia
supernovae extremely bright optical events. In addition, observations
indicate that Type Ia supernova	light curves form a homogeneous	class
with relatively	small intrinsic	scatter	\citep{filippenko97}.  The
brightness and homogeneity of the light	curves have earned these
supernovae ``standard candle" status for determining distances in the
Universe and thus cosmological parameters \citep[][and references
therein]{sandage+93,branch98,garnavich+98,perlmutter+99,turner+01,tonry+03,knop+03}.
Despite	the apparent homogeneity of these events, some observations
show exceptions	\citep{li+01}.

In the classic Type Ia scenario, a carbon flash	evolves	into a
large-scale explosion, eventually consuming the	entire white dwarf.
The thermonuclear flame	propagates through the C-O fuel	as either a
subsonic deflagration front
\citep{nomoto+76,nomoto+84,reinecke+02,gamezo+03} or a supersonic
detonation wave	\citep{arnett69,boisseau+96} and releases sufficient
energy to unbind the star.  Models involving either a pure
deflagration or	a pure detonation, however, are	unable to provide an
explanation for	both the observed expansion velocities and the
spectrum produced by ejecta that are rich in intermediate-mass and
iron-peak elements \citep{truran+71,woosley+86,woosley+90}.

Considerable effort has	gone into resolving the	discrepancy with
observations. One possibility is a transition of the initial deflagration
front into a detonation	as the burning front propagates	outward
through	the outer layers of the	white
dwarf~\citep{blinnikov+87,khokhlov91-dd,niemeyer+97}.  Another
possibility is that the	initial	deflagration dies out as a result of
the expansion of the outer layers of the white dwarf.  When these
gravitationally-bound layers recollapse	onto the white dwarf, a
detonation ensues, incinerating	the remaining nuclear
fuel~\citep{khokhlov91-pd}.  These models are often referred to	as
``delayed detonation'' models. In addition, our	knowledge regarding
initial	conditions characterizing the stellar core just	prior to
ignition is severely limited. Analytic models suggest rather complex
flow patterns \citep{woosley+03}, while	the first multidimensional
hydrodynamic models of stellar interiors became	available only
recently \citep{hoeflich+02}.

Despite	these deficiencies, delayed detonation models of massive C-O
rich white dwarfs are capable of accounting for	most Type Ia
characteristics, including the observed	expansion velocities of
silicon-group and iron-group nuclei.  In this Letter we report
on research into the initial deflagration phase	of the explosion and
present	the results of simulations in which ignition occurs slightly
off-center.

\section{Numerical model}

Modeling complex events	like Type Ia supernovae	is limited by
available computing resources due to the vast range of length scales
involved.  Theoretical models of thermonuclear flame fronts indicate
that, for conditions typical of	white dwarf interiors, the flame
thickness is between 8 and 12 orders of	magnitude smaller than the
stellar	radius \citep{timmes+92}.  Because of this difference in
length scales, multidimensional	Type Ia	models must make use of	an
appropriate subgrid model for the evolution of the thermonuclear
burning	front.

A subsonic burning front originating near the center of	a massive
white dwarf is subject to several fluid
instabilities~\citep{niemeyer+kerstein97}. These instabilities,
dominated by the Rayleigh-Taylor instability (RTI) on the largest
scales,	increase the surface area of the flame and its effective
propagation speed.  As was proposed by \cite{khokhlov95},	this
increase in speed is offset by the effects of the RTI: the more
quickly	the flame advances, the	more rapidly the bubbles produced by
RTI merge.  The	surface	area and flame speed, therefore, do not
increase.  Khokhlov's self-regulating mechanism	was observed in
numerical simulations of compressible as well as incompressible	flows
\citep{khokhlov95,vladimirova+03}, and current state-of-the-art
numerical models \citep{reinecke+02,gamezo+03} indicate	that this
mechanism might	be robust.  We will report in an upcoming publication
a series of simulations	that confirm these results.

In our modeling	effort,	we performed entire-star simulations of	the
deflagration phase of Type Ia supernovae using the FLASH code, a
parallel adaptive-mesh multi-physics astrophysical hydrodynamics code
\citep{fryxell+00,calder+02}.  The FLASH code solves the Euler
equations for compressible flow	and the	Poisson	equation for
self-gravity.  
A recent addition is a custom implementation of the flame capturing
scheme of \cite{khokhlov01} in which the flame advances by evolving a
passive scalar variable with an advection-reaction-diffusion equation.
Reaction and diffusion coefficients are chosen to keep the interface
several grid points wide and to propagate with the given
speed. Specifically, we take the flame speed to be a maximum of a
laminar flame speed and \cite{timmes+92} and a turbulent flame speed,
based on the assumption that the turbulent burning on a macroscopic
scales is driven by RTI \cite{khokhlov95} and described above. The
flame model controls the transition from carbon to magnesium and
produces the corresponding amount of energy.  Further transition to
silicon and nickel are modelled as two consecutive stages of
distributed burning with exponential decays on timescales that
depend on temperature and density.

For our	initial	configuration, we adopt	a ``cold'' isothermal model of
a white	dwarf with a temperature of $T_\mathrm{wd}=5\times10^7$~K, a
mass $M_\mathrm{wd}=1.36 M_\sun$, and a	radius
$R_\mathrm{wd}=2.13\times10^8$~cm.  We found that this model was not
in hydrostatic equilibrium after being interpolated onto the FLASH
simulation mesh	due to a small mismatch	in discretization of the
solution and slight differences	in the equation	of state between that
of the initial analytic	model and that used in the simulation. To
bring the model	into hydrostatic equilibrium, and therefore ensure
that the subsequent evolution is not affected by this discrepancy, we
used a variant of the relaxation method	proposed by
\citet{arnett94}. The remaining	mismatch in the	relaxed	model produced
velocities smaller than	10~km\,s$^{-1}$	during test simulations
executed for several seconds of	evolution time.	 We will present these
and other verification tests of	the method in a	forthcoming
publication~\citep{calder+04}.

For the	multidimensional simulations, we mapped	the relaxed model onto
the adaptive simulation	mesh generated according to the	following
refinement criteria.  We forced	the resolution in the innermost	part
of the grid containing the bulk	of the star, $r_\mathrm{uni} = 2300$
km, to be no worse than	a predefined limit, $\Delta r_\mathrm{uni} =
12.8$~km.  During the simulation, the same resolution was also
enforced in any	grid region where the density exceeded
$\rho_\mathrm{uni} = 3\times10^6$~g\,cm$^{-3}$.	Grid refinement	was
also active in the region where	local density variations were greater
than 20\%.  We used the	same criteria to resolve velocity variations
when the velocity exceeds 100~km\,s$^{-1}$.  Finally, we resolved the
region occupied	by the flame front identified by the passive scalar at
the highest resolution.	 These refinement criteria result in an
initial	grid configuration that	uniformly resolves the interior	of the
star, and the resolution gradually decreases with radius outside of
the star.  During the evolution, the mesh adapts so that only the
regions	at high	densities are refined.

We initialized the flame front by placing a small spherical region of
completely burned material in hydrostatic equilibrium with its
surroundings near the stellar center. We note that our configuration
differs	significantly from an off-center ignition model	considered
by~\cite{niemeyer+96} in which the ignition region was placed much
farther	from the stellar center.  We determined	the subsequent
evolution of the flame by integrating the hydrodynamic equations
supplemented by	the advection-diffusion-reaction equation.  We used a
tabular	equation of state suitable for the dense stellar material and
performed the simulation in Cartesian geometry in three	dimensions. At
the boundaries of the domain, we imposed hydrostatic outflow-only
boundary conditions.  We calculated the	gravitational field assuming
spherical symmetry, an assumption justified by the relatively small
displacement of	mass during the	simulation.

\section{Results}

In the first simulation, we used a computational domain	with sides $L
= 6.5536 \times	10^8$~cm and an	effective resolution of	1024$^3$
(6.4~km); in the second	simulation, we used the	same domain but	with
an effective resolution	of 4096$^3$ (1.6~km).  Due to the large	size
of the high-resolution simulation, the evolution after $t = 0.76$~s
was calculated at an effective resolution of 3.2~km. We initiated the 
nuclear
flame in a spherical region of radius of 50~km centered	at $(x,y,z) =
(10,6,4)$~km (i.e., 12~km off-center). This displacement corresponds
to about 0.25 of the radius of the ignition region, a few percent of
the expected radius of the convective region in	the center of the
progenitor, and	$6 \times 10^{-3}$ of the radius of the	star. This is
significantly smaller than the 200~km displacement used	by
\citet{niemeyer+96}.  Asymmetric placement of the ignition region was
aimed at minimizing possible mesh discretization effects on the	early
flame evolution. (Verification tests with a centrally ignited flame
indicated that mesh effects are	strong enough to promote preferential
instability growth along the grid axes.)

\begin{figure}[ht]
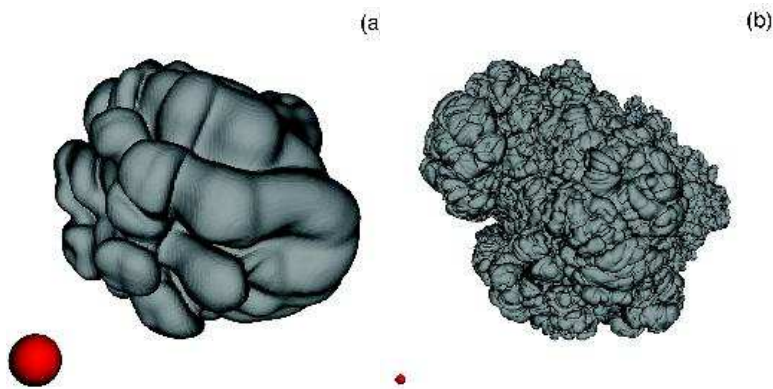

\begin{center}
\resizebox{5cm}{!}{\includegraphics{f1a.epsi}}
\resizebox{5cm}{!}{\includegraphics{f1b.epsi}}
\end{center}
\caption{Flame surface evolution.
a) The bubble at the moment the
Rayleigh-Taylor	unstable top surface begins to grow rapidly (i.e, at
the end	of the first phase of its evolution).
b) The bubble as it approaches the stellar surface (i.e., at
the end	of the second phase of its evolution).
The complex structure results from the merging of secondary and	later
generations of bubbles produced	by RTI.	Filaments and sheets of
largely	unburned material are present throughout the bubble. For
comparison, the	red sphere has a radius	of 25~km in both images.
}
\label{f:bubble_evolution}
\end{figure}

\begin{figure}[ht]
\begin{center}
\resizebox{5cm}{!}{\includegraphics{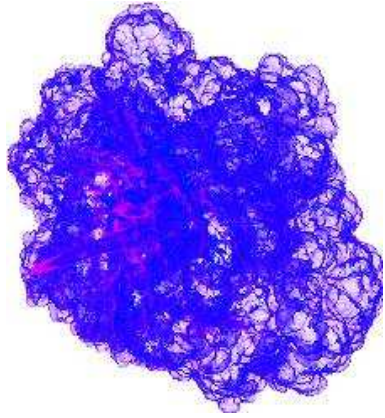}}
\end{center}
\caption{Volume	rendering of the bubble	of hot ash as it approaches
the stellar surface (i.e., at the end of the second phase of its
evolution).
The complex structure is formed	by the merging
of secondary and later generations of bubbles produced by RTI.
The length scale in the	image is identical to that of 
Fig.~\ref{f:bubble_evolution}(b).
}
\label{f:bubble_structure}
\end{figure}

\begin{figure}[t]
\begin{center}
\includegraphics[width=2.7truein,clip=true,angle=0]{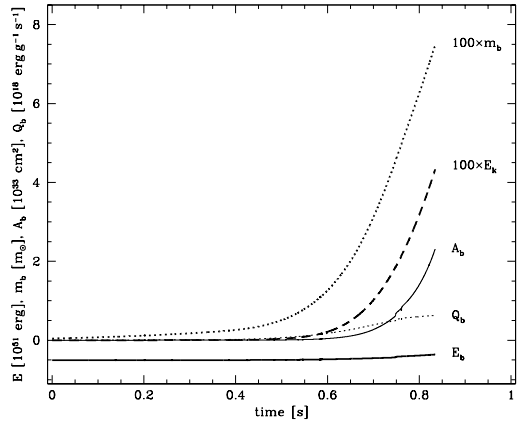}
\end{center}
\caption{
Some global properties of the model as a function of time. Shown are
the total binding energy $E_\mathrm{b}$	(thick solid), the energy
generation rate	$Q_\mathrm{b}$ (dotted), the flame front area
$A_\mathrm{b}$ (thin), the total kinetic energy	$E_\mathrm{k}$ (thick
dashed), and the mass of the bubble $m_\mathrm{b}$ (thick dotted).
}
\label{f:bubble_properties}
\end{figure}

We find	that the evolution of the flame	front (the bubble of hot ash)
consists of three distinct phases.  The	first lasts for	approximately
0.3--0.4 s and is a period of slow, mostly radial growth of the
bubble.	The position of	the hot	bubble changes only slightly, but one
can observe a slow divergence of its shape from	spherical symmetry,
caused by the non-uniformity of	the gravitational field.  The next
phase is characterized by substantial growth of	the asymmetry created
during the first phase and the development of RTI at the bubble
surface. At this point,	the difference between the turbulent flame
speed at the ```top'' (further away from the center of the star) and
the ``bottom'' (closer to the center) of the bubble leads to further
distortion of the bubble. Later	in this	phase, the top of the bubble
becomes	highly susceptible to RTI, which leads to the successive
formation of a series of effusive plumules (see
Fig.~\ref{f:bubble_evolution}(a)). This structure rapidly grows in size
and complexity (see Fig.~\ref{f:bubble_evolution}(b) and
Fig.~\ref{f:bubble_structure}),	increasing the surface area of the
flame and therefore the	burning	rate and rate of energy	release	(see
Fig. ~\ref{f:bubble_properties}). The whole structure starts to
rapidly	ascend toward the surface.  As in the first phase, the bubble
does not encounter significant density gradients, and the evolution of
the surface is dominated by the	flame evolution.

\clearpage

\clearpage

The third phase	of the evolution is marked by the time at which	the
bubble reaches the surface layers of the star where there is a large
decreasing density gradient, and lasts for only	$\approx0.05$~s.  Here
pressure effects become	important to the evolution, and	lateral
expansion accompanies the forward motion. The evolution	is extremely
fast.  At $0.78$~s, the	tip of the bubble is located at
$r\approx 1.51\times10^{8}$~cm and by $t = 0.83$~s the bubble
reaches	a radius of $\approx 1.88\times10^{8}$~cm.  By the end of the
simulation, the	bubble has a radial extent of about 1500~km and	a similar
diameter. The interior of the bubble has a highly complex structure,
with a tangled network of burned and partially burned material having
bulk velocities	approaching 10,000~km\,s$^{-1}$	and lateral expansion
in the range of	3000~km\,s$^{-1}$. The rapid expansion of
the bubble and the associated decrease of density and temperature
result in a slowing of the thermonuclear reaction rates	and flame
extinction.  The physics expected in the subsequent evolution are not
properly accounted for in our model (any burning will no longer	have
the form of a well-defined deflagration	front),	so at this point we
ended the simulations.

Figure~\ref{f:bubble_properties} depicts temporal evolution of select
global quantities.  The	result is that $\approx	0.026 M_\sun$
($\approx2\%$ of the stellar mass) burns into iron-peak	elements.  The
rapidly	rising bubble expels burned material originating in the	center
of the star with a characteristic velocity of 5000~km\,s$^{-1}$	and
with local velocities in excess	of 10,000~km\,s$^{-1}$.	Only about
0.2\% of the stellar mass has sufficient energy	to leave the system,
though.	 The rest of the burned	material remains gravitationally bound
and will fall back onto	the surface of the star, thereby increasing
the abundance of Ni-group elements in the surface layers.

\section{Discussion and	Conclusions}

In this	paper, we studied the hydrodynamics of thermonuclear
deflagration of	a massive white	dwarf. We considered a flame ignited
slightly off-center and	followed its evolution in a full
three-dimensional simulation of	an entire white	dwarf with no imposed
symmetry.  We observed the formation of	a Rayleigh-Taylor-unstable
bubble of hot nuclear ash and its rapid	rise toward the	stellar
surface.  The rising bubble takes the form of a	typical
mushroom-shaped	plume and is composed of intermediate and iron-peak
elements. As it	rises, the bubble's motion is continuously powered by
buoyancy due to	heating	from the energy	released by nuclear burning,
and supersonic speeds are achieved. Further, the morphology of the
surface	of the bubble, determined by RTI, displays characteristics
similar	to those obtained in earlier numerical studies of deflagrating
white dwarfs.  In particular, generations of merging bubbles form on
the surface of the initial bubble.

Although this model does not produce an	immediate explosion, it
provides a robust mechanism for	the transportation of heavy elements
to the stellar surface.	 If combined with a successful explosion, such
a surface composition might offer an explanation for the presence of
iron-group elements at high velocities,	as observed in SN 1991T-like
supernovae \citep{filippenko97,jeffery+92}. Our	model produces
$\approx 0.03 M_\sun$ of heavy elements	at the stellar surface without
invoking an ad hoc transition to detonation \citep{yamaoka+92}.

Since the buoyancy-driven velocity created soon	after the ignition
quickly	becomes	comparable to the expected velocity of the convective
motions	\citep{hoeflich+02}, convection	is unlikely to influence the
bubble's evolution in any significant way. This	conclusion is
consistent with	the results of \cite{niemeyer+96}.

The fact that our initial conditions deviate from perfect symmetry by
only a minute amount, but lead to a result drastically different from
a central ignition model, indicates that central ignition models in
general	are unlikely to	be typical initial conditions for the
explosion. Our results therefore strongly disfavor the classic
central-ignition pure deflagration scenario.

In future studies, we plan to focus on the late	time evolution of this
model~\citep{plewa+04}.	Recollapse and/or detonation are likely
outcomes.  Detailed calculations of the	nucleosynthetic	yields and
further	convergence studies will be required to	provide	a quantitative
link between this model	and observations.  Finally, it will be
important to study the evolution of the	deflagration front in the
presence of strong shear, which	can influence the evolution of the
rising bubble, in more detail.

\acknowledgements

The authors thank Peter	H\"oflich and Alexei Khokhlov for their
comments. The authors also acknowledge visualization support from the
ANL Futures Laboratory.	This work is supported in part by the
U.S. Department	of Energy under	Grant No. B341495 to the Center	for
Astrophysical Thermonuclear Flashes at the University of Chicago.

\end{document}